# Performance Comparison of Two Streaming Data Clustering Algorithms

Chandrakant Mahobiya[1] , Dr. M. Kumar[2]
[1,2]Department of Computer Science & Engineering
Sagar Institute of Research & Technology, Bhopal, India

*Abstract*—The weighted fuzzy c-mean clustering algorithm (WFCM) and weighted fuzzy c-mean-adaptive cluster number (WFCM-AC) are extension of traditional fuzzy c-mean algorithm to stream data clustering algorithm. Clusters in WFCM are generated by renewing the centers of weighted cluster by iteration. On the other hand, WFCM-AC generates clusters by applying WFCM on the data & selecting best K± initialize center. In this paper we have compared these two methods using KDD-CUP'99 data set. We have compared these algorithms with respect to number of valid clusters, computational time and mean standard error.

## I. INTRODUCTION

Recently, a huge amount of streaming data, such as network flows, phone records, sensor data, and web click streams have been generated because of progress in hardware and software technologies. Analyzing these data has been a hot research topic [1], [2], [3], [4], [5]. Data streams are temporally ordered, fast changing, massive, and infinite sequence of data objects [6]. Clustering streaming data is becoming important due to the availability of large amount of data recorded from various sources. Clustering is a division of data into groups of similar objects. Each group is called a "cluster" and contains objects that are similar between them and dissimilar compared to objects of other groups. Streaming data flow continuously for days, months, or even years. It might not be possible to store all the data in memory, but necessary to analyze and delete it. Streaming data may have the limitation that it cannot be revisited and has to be processed as it comes, that is, no random access is possible. Finding meaningful clusters under these limitations is challenging. A number of algorithms have been proposed recently which cluster streaming data by using a single pass approach [7]. Many traditional approaches are also extended for streaming data, FCM is one of them. Fuzzy c-means (FCM) is a method of clustering in which a data point can assign to more than one cluster at the same time. The FCM clusters total data set, and data stream may contain large data sets, so FCM may consume significant amounts of CPU time to cluster. To overcome this problem FCM is extended to weighted fuzzy c-mean (WFCM) & weighted fuzzy c-mean adaptive cluster number (WFCM-AC) for data stream.

In this paper we have made an attempt to compare the performance of WFCM algorithm and WFCM-AC algorithm in terms of number of valid clusters, error rate & computational time

The rest of the paper is organized as follows: In section 2 we have surveyed related works. FCM and WFCM approach are explained in section 3. Comparative results are discussed in section 4. Finally conclusion and comparative results are discussed.

## II. LITERATURE SURVEY

BIRCH can be considered as a primitive method of clustering data streams [8]. In fact it has been designed for traditional data mining and is not suitable for very large data sets like data streams. This method introduces two new concepts: micro clustering and macro clustering. The first well-known algorithm performing clustering over entire data streams is the STREAM algorithm proposed by Guha et al. [9], [10]. The STREAM algorithm extends the classical k-median in a divide-and-conquer fashion to cluster data streams in a single pass. Babcock et al. [11] proposed to extend the STREAM algorithm from one-pass clustering to the sliding window model, where data elements arrive in a stream and only the last N elements are considered relevant at any moment. The CluStream framework proposed in [12] is effective in handling evolving data streams. It divides the clustering process into an online component which periodically uses micro clusters to store detailed summary statistics and an offline component which uses this summary statistics in conjunction with other user input to produce clusters. For high-dimensional data stream clustering Aggarwal et al. [13] proposed HPStream, which reduces the dimensionality of the data stream via data projection before clustering. Cao et al. [14] proposed a DenStream algorithm, which extends DBSCAN by introducing micro clusters to the density-based connectivity search. Chen and Tu [15] also proposed a density-based method termed D-Stream. They mapped new data points into the corresponding grid to store density information, which was utilized for clustering. Khalilian et al. [16] have improved K-Means method by using divide and conquer method. Experimental results show that it is capable to cluster objects in high quality and efficiency especially in objects with high dimensional. Recently [17] have developed a connectivity based reprehensive points to cluster data stream. Accuracy is outstanding in their research but it exhibits low performance. Another point is using a repository for previous data so it is unable to give us a history in different scale time. E-Stream [18] is a data stream clustering technique which supports following five type of evolution in streaming data: Appearance of new cluster, Disappearance of an old cluster, Split of a large cluster, merging of two similar clusters and change in the behavior of cluster itself. It uses a fading cluster structure with histogram to approximate the streaming data. Though its performance is better than HPStream algorithm but it requires many parameters to be specified by user. DUCstream (Dense Units Clustering for data stream) is proposed by Gao. et al. [19]. It is a grid based technique which divides the data space in the non-overlapping grids and process data in the form of chunks. Chen et. al have presented DD-Stream[20]. It combines the density based and grid based clustering approaches. HUE-Stream [21] extends E-Stream which is described earlier, in order to support uncertainty in heterogeneous data. In [22], a multistage random sampling method is proposed to speed up fuzzy c means. There are two phases in the method. In the first phase, random sampling is used to obtain an estimate of centroids and then fuzzy c means (FCM) is run on the full data with these initialized centroids. In [23], speeding up is obtained by taking a random sample of the data and clustering it. The centroids obtained then are used to initialize the entire data set. Richards and James [24] proposed a sampling based method for extending fuzzy and probabilistic clustering to large or very large data sets. In [25] an algorithm called "AFCM" is used to enhance FCM in terms of speed. This is done using lookup table. In [26], authors proposed number of efficient and scalable parallel algorithms for a special purpose architecture description of a modified FCM algorithm known as 2rFCM is given. A fast FCM algorithm is proposed in [27]. They have used concept of decreasing the number of distance calculations by checking the membership value for each point.

### III. COMPARISON METHODOLOGY

We first give brief description of two algorithms:

*A. WFCM algorithm*

The concept of weighted Fuzzy clustering Algorithm was introduced by R. Wan et al. [28 z]. The WFCM is based on the concept of renewing the weighted clustering centers using iterations till the cost function gets a satisfying result or the number of iteration is equal to a tolerance. The steps involved in WFCM are as follows: first data are divided into chunks according to the reaching time of data. The size of each chunk is determined by main memory of the processing system. Since data are continuous, a time weight $w(t)$ is applied on each data representing the data's influence extent on the clustering process. Data numbers are represented by $n1$, $n2$,…, $ns$ of chunks $X1$, $X2$,……, $Xs$. Chunks of data are imported in FCM algorithm to get cluster centroids and after that the procedure computes centroid weight $w_i$ by summing the multiplication of $u_{ij}$ & $w_j$ then $w_i$ updated. In next step WFCM updated cluster centroids and objective function are calculated. Algorithm stops when objective function is minimized or concentrates on a certain value or its improvement over previous iteration is below a certain threshold or iterations reach a certain tolerance value. We then compute new membership matrix U and the procedure switch in to weight update step if l=s else all the steps are executed again.

*B. WFCM-AC algorithm*

Mostafvi and Amiri [29] extended WFCM algorithm and called it. In WFCM-AC initially FCM is applied to the normalized chunk of stream data, that each data point is replaced by subtracting the total mean and dividing the result by the standard deviation. Membership value of each point is represented by membership matrix [29]. Each obtained center can be weighted by summing the membership values of all examples that have partial belonging to it. All points of processed chunk are discarded and the obtained centers are used for clustering the next chunk. To use the information from past history, the data points in the newly arrived chunk are clustered with the weighted centers of the last chunk. Because of evolving feature of stream data, the number of clusters can change over time. To adapt to this change, WFCM-AC uses the advantage of slow change of clustering structure in data stream. All data that should be clustered are also partitioned with k± clusters, where k is the number of current clusters. To increase the number of clusters by one, each data point is assigned to the cluster in which it has the highest membership value. The farthest point in each cluster is chosen as new center. Whenever a new cluster center is found, then WFCM algorithm is applied to all current data with the k+1 initialized centers. This procedure is repeated k times. The quality of clustering structure is measured for each run of the algorithm and the best structure is then taken. To decrease the number of clusters by one, each time one of the seed points is temporarily eliminated and WFCM is applied to all current data with the k-1 remaining initialized centers. Finally the structure that improves the quality measure will be chosen.

### IV. EXPERIMENTAL SETUP AND RESULTS

For execution of two algorithms, we have used MATLAB 7.8.0 tool, operating system-Window-7(32 bit), CPU-2.40 GHz, RAM- 2GB and hard disk of 500 GB. We evaluated performance of WFCM and WFCM-AC clustering algorithms using KDD'99 attack datasets. In KDD99 dataset these four attack classes (DoS, U2R, R2L, and probe) are divided into 22 different attack classes. The 1999 KDD datasets are divided into two parts: the training dataset and the testing dataset. The testing dataset contains not only known attacks from the training data but also unknown attacks. Since 1999, KDD'99 has been the most wildly used data set.

The cluster formation using WFCM algorithm is shown in the Fig 1.1 and WFCM-AC is shown in the Fig 1.2. The KDD'99 normal data is represented by *, Dos by *, Prob by *, U2R *respectively. The Fig 1.3 shows number of valid clusters generated versus number of chunks of stream data. The two algorithms generates the same number of clusters. The formation of clusters gives the information of valid and invalid cluster according to cluster valid index.

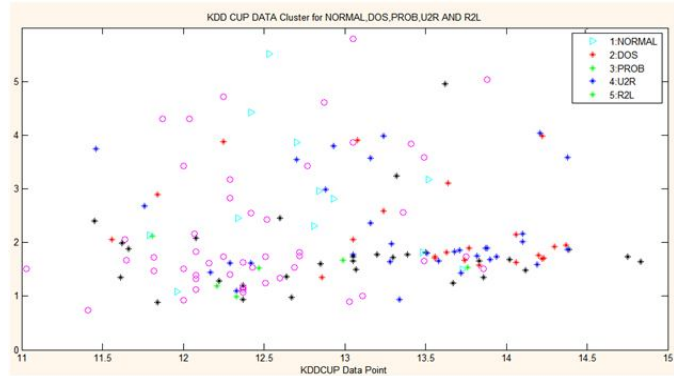

Fig 1.1 Generated clusters of KDDCUP99 data point WFCM.

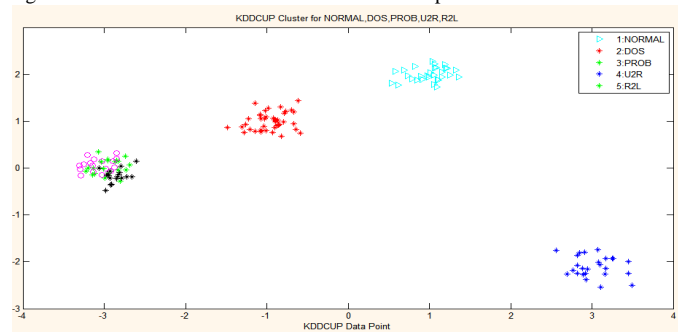

Fig 1.2 Generated clusters of KDDCUP99 data point by WFCM-AC.

The evaluation of clustering performance used some standard parameter such as number of valid cluster generation and number of cluster along with mean absolute error of clustering process. The mean absolute error process induced the error rate of clustering technique. The process of clustering used data chunks of sizes 1000, 2000, 3000 and 4000 (small data to large size of data). For measurement of error in cluster formation, we have used standard formula given below in equation (1). In clustering the mean absolute error (MAE) is a quantity used to measure how close real or predictions are to the eventual outcomes. The mean absolute error is given by:

$$MAE = \frac{1}{n}\sum_{i=1}^{n}|f_i - y_i| = \frac{1}{n}\sum_{i=1}^{n}|e_i| \qquad (1)$$

As the name suggests, the mean absolute error is an average of the absolute errors $e_i = |f_i - y_i|$, where $f_i$ is the prediction and $y_i$ the true value.

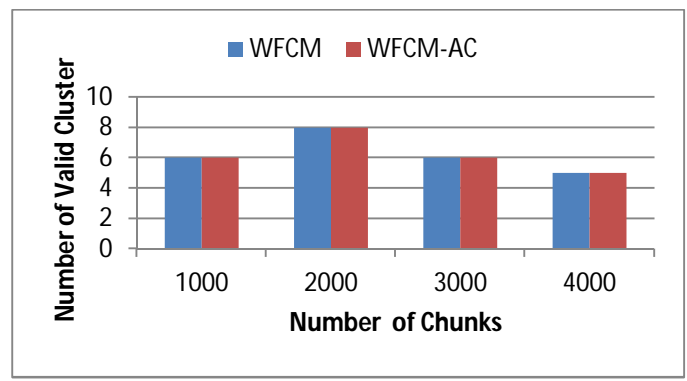

Fig. 1.3 Number of valid clusters generated according to size of data.

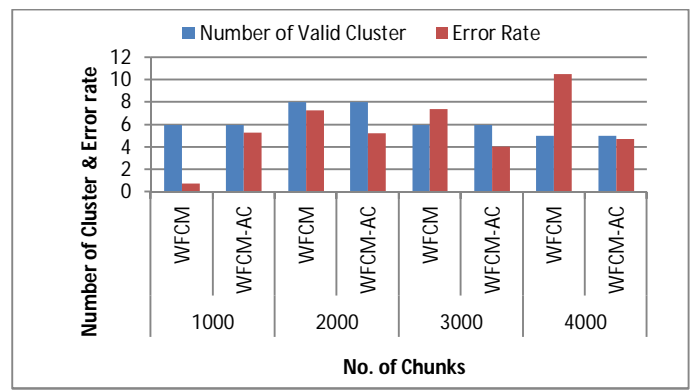

Fig. 1.4 Number of valid clusters & MAE for WFCM and WFCM-AC algorithm.

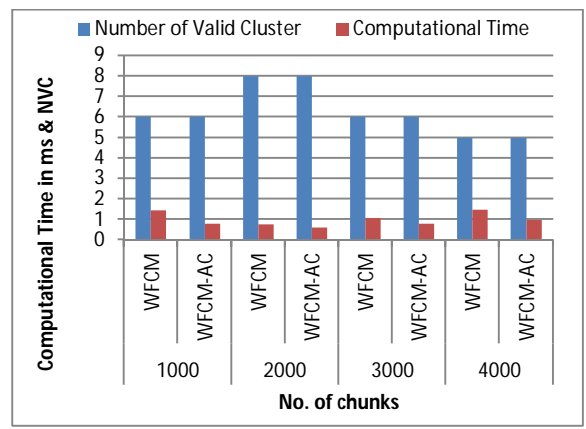

Fig. 1.5 Comparative result graph for Iteration time between the method WFCM and WFCM-AC.

Figure 1.3 shows number of valid clusters and the corresponding mean absolute error for WFCM & WFCM-AC. The Figure 1.4 shows the number of valid clusters and the iteration time taken by WFCM & WFCM-AC.

## V. CONCLUSION

In this research, we have compared performance of two algorithms, viz., WFCM-WFCM-AC. It should be noted that the number of valid clusters are same for various chunks of data streams (Fig 1.3). The mean absolute error (MAE) for WFCM is low compared to WFCM-AC for 1000 data chunks, but as the size of data chunks increase the performance of WFCM-AC is better as low MAE compared to WFCM. Further it may be observed that iteration time for the formation of valid clusters is always less for all data chunk sizes (Fig 1.4) for WFCM-AC compared to WFCM. It can be said that WFCM-AC performed better on all fronts and should be preferred compared to WFCM.